\documentclass[11pt,letter,reqno]{amsart}
\usepackage{tabmac}	
\usepackage{amsmath}
\usepackage[active]{srcltx}						
\usepackage{amssymb}										
\usepackage{braket}											
\usepackage{amsfonts}										
\usepackage{amssymb}										
\usepackage{mathrsfs}										
\usepackage{tabmac}
\usepackage{latexsym}
\usepackage{amsthm,amsopn,stmaryrd}
\usepackage[active]{srcltx}	
\usepackage{graphicx}										
\usepackage[usenames,dvipsnames,svgnames,table]{xcolor} 	
\usepackage[breaklinks, hypertexnames=false]{hyperref}					
\usepackage[utf8]{inputenc}
\usepackage{graphicx}																	
\usepackage{pstricks}																	
\usepackage{pst-plot}																	
\usepackage{pst-math} 																	
\usepackage{pstricks-add}																
\usepackage{pst-func}																	
\usepackage{pst-optic}	
\usepackage[tiling]{pst-fill}
\usepackage{fp}

\def\Square{\pspicture(0.5,0.5)\psframe[dimen=middle](0.5,0.5)\endpspicture}

\newcommand{\tcercle}[1]{\ensuremath{\setlength{\unitlength}{1ex}\begin{picture}(2.8,2.8)\put(1.4,1.4){\circle{2.7}\makebox(-5.6,0){#1}}\end{picture}}}

\def\AR{{\square^\mathrm{R}_{r,s}}}

\let\La\Lambda
\let\Om\Omega

\let\ta\theta

\let\rw\rightarrow
\let\xrw\xrightarrow
\let\ga\gamma

\def\R{{\rangle}}

\let\ti\tilde
\let\a\alpha
\let\d\partial

\newcommand{\LL}{\ensuremath{\langle\!\langle}}
\newcommand{\RR}{\ensuremath{\rangle\!\rangle}}
\def\lrw{\leftrightarrow}

\def\cb#1{{\color{blue}#1}}

\def\cd{{\circledast}}
\def\tr{{\tilde r}}
\def\ts{{\tilde s}}
\def\beq{\begin{equation}}
\def\eeq{\end{equation}}
\def\bea{\begin{align}}
\def\eea{\end{align}}
\def\up{{\uparrow}}
\def\dn{{\downarrow}}

\begin{document}


\title[Ramond singular vectors]{Ramond singular vectors and Jack superpolynomials}

\author{Ludovic Alarie-Vezina}
\address{D\'epartement de physique, de g\'enie physique et
d'optique, Universit\'e Laval,  Qu\'ebec, Canada,  G1V 0A6.}
\email{ludovic.alarie-vezina.1@ulaval.ca}

\author{Patrick Desrosiers}
\address{Instituto de Matem\'atica y F\'{\i}sica, Universidad de
Talca, Casilla 747, Talca, Chile.}
\email{patrick.desrosiers@inst-mat.utalca.cl}

\author{Pierre Mathieu}
\address{D\'epartement de physique, de g\'enie physique et
d'optique, Universit\'e Laval,  Qu\'ebec, Canada,  G1V 0A6.}
\email{pmathieu@phy.ulaval.ca}

\thanks{The authors thank Luc Lapointe for useful discussions.  The work of P.M. was supported by NSERC.   The work of P.D. was supported by CONICYT through FONDECYT grant  \#1131098 and  Anillo de Investigaci\'on ACT56.}

   \begin{abstract}
    The explicit formula for the superconformal singular vectors in the Neveu-Schwarz sector has been obtained recently, via its symmetric polynomial representation, as a sum of
Jack superpolynomials.  Here we present the analogous, but slightly more complicated, 
 closed-form expression for the Ramond singular vectors.
\end{abstract}


 \maketitle

\setcounter{page}{0}
	\pagenumbering{arabic}
	
\small
\normalsize



	\section{Introduction}

	The remarkable  correspondence between the singular vector in Virasoro $(r,s)$--Kac module  and the Jack polynomial indexed by the rectangular diagram with partition $(r^s)$ \cite{MY,SSAFR,AMOSa}  has been extended to the superconformal case in \cite{DLM_jhep}.   In that context, singular vectors are represented by sums of Jack superpolynomials (sJack). The main difference between the two cases is thus that the one-to-one correspondence of the former situation  is lost in the latter one.
\footnote{In that regard, we should point out the amazing observation made in \cite{BBT} which is that a superconformal singular vector can be represented by a single (rank 2) Uglov polynomial \cite{Uglov}. These polynomials  are the specialization of the Macdonald polynomials \cite{Mac}  at $q=t=-1$. Notice that these do not involve any anticommuting variables. As it will be reviewed below, the polynomial representation of states is obtained via a free-field representation. In the superconformal case, the algebra generators are expressed in terms of a free fermion and a free boson. The Uglov-polynomial representation follows once the
free fermion is bosonized \cite{CFT}. 
Note that in this construction, the differential-operator representation of the super-Virasoro generators, 
as well as the one-to-one correspondence between states and  symmetric superpolynomials, are both lost. Understanding the connection between the Uglov polynomials and the sJacks is a puzzling issue.}\\

However, the rectangle rule still holds in disguise: the contributing terms are indexed by superpartitions which are self-complementary (in a sense which differs slightly between the Neveu-Schwarz (NS) and the Ramond (R) sectors) and such that the superpartition glued to its $\pi$-rotated version (and slightly modified in the NS sector) fills a rectangle with $r$ columns and $s$ rows. This is a severe constraint on the allowed terms. The example presented in \cite{DLM_jhep} illustrating the strength of this restriction is the NS singular vector at level 33/2 which has 11 terms, while there are 1687 states at that level (and thus, the same number of superpolynomials in a state representation expanded in a generic basis).\\

For the NS sector, a closed-form expression in terms of sJacks has been displayed in \cite{DLM_jhep}. However, we failed to obtain a similar result in the R case. Indeed, the latter is much more difficult to cope with. To illustrate the additional level of complexity, we simply note that the highest-level R singular vector that we have generated is for $(r,s)=(5,6)$, hence at level 15 (thus a level lower than the previous NS example), and it contains 86 contributing terms (out of a total of 1472 possible states). This signals a radical increase in complexity.  It can be traced back to the R version of the rectangle condition, which is weaker that in the NS case and allows thus for more terms. \\

But, somewhat surprisingly, a closed-form expression has actually been found. Its exposition is the aim of this work.\footnote{We recall that until the publication of \cite{DLM_jhep,BBT}, the only singular vectors with known closed-form expressions   were those   with either $r=1$ or $s=2$ (see \cite{BsA} and \cite{Watts} for the NS and R sectors, respectively).} It is presented in Section 3. The formula has been tested up to level $rs/2=15$, inclusively.  The next section is a review of relevant results from \cite{DLM_jhep}. The required background on superpartitions and sJacks is collected in Appendix A.\\

\section{Fundamental Correspondences}


\subsection{Representation of Ramond states as symmetric superpolynomials}

Let us first review the connection between symmetric superpolynomials and states in R highest-weight modules. Recall that in the R sector of the super-Virasoro algebra \cite{BKT,FQS,Dorr}\begin{align}\label{svir}
&[L_n,L_m]=(n-m)L_{n+m}+\frac{c}{12}n(n^2-1)\delta_{n+m,0}\nonumber\\
&[L_n,G_k]=\left(\frac{n}{2}-k\right)G_{n+k}\nonumber\\
&\{G_k,G_l\}=2 L_{k+l}+\left(k^2-\frac{1}{4}\right)\frac{c}{3}\delta_{k+l,0},
\end{align}
 $G$ indices are integers: $k,l \in\mathbb{Z}$.    We  denote by $|h\R$ the positive-chirality\footnote{In other words, 
$|h\R\equiv |h\R^+$ and the negative chirality-sector is built from the highest-weight state $|h\R^-=G_0|h\R^+$. } highest-weight vector of conformal dimension $h$, i.e., 
\beq L_{n}|h\R=0=G_n|h\R\quad \forall n>0\qquad \text{and}\qquad L_{0}|h\R=h|h\R.
\eeq The highest-weight module $\mathscr{M}$ is generated by all states of the form
\begin{equation} G_{ -\La_1 }\cdots G_{ -\La_m }L_{-\La_{m+1}}\cdots L_{-\La_\ell} |h\R
\end{equation}
with the conditions
\begin{equation} \label{eqspart}  \La_1>\ldots> \La_m\geq 0,\qquad \La_{m+1}\geq \ldots \geq \La_{\ell}> 0,
\qquad \text{with}\qquad m\geq 0,\qquad \ell \geq 0.\end{equation}
 The last equation simply means that $\La=(\La_1,\ldots,\La_m;\La_{m+1},\ldots,\La_\ell)$ is a superpartition of fermionic degree $m$ (cf. Section \ref{A1}).

The state $|\chi\R$  is a singular vector if $G_n|\chi\R=0$ and $L_n|\chi\R=0$ for all $n>0$.   All these constraints are consequences  of the following two conditions:  
\begin{equation}  \label{condsingvecR} G_1|\chi\R=0\qquad\text{and}\qquad L_1|\chi\R=0.
\end{equation}
To explore singular vectors, it is sufficient to focus on Kac modules (e.g., see \cite{Dorr}). Recall that a R highest-weight module  is a  Kac module  whenever  the central charge $c$ 
and the conformal dimension $h$ are related via the parametrization:
\begin{equation} \label{hrsR} 
c=\frac{15}{2}-3\left(t+\frac{1}{t}\right)\qquad\text{and}\qquad h_{r,s}=\frac{t}{8}(r^2-1)+\frac{1}{8t}(s^2-1)-\frac{1}{4}(rs-1)+\frac{1}{16},
\end{equation}
where $r$ and $s$ are positive integers such that $r-s$ is odd, while $t$ is a complex number. 
In such a module, there is a singular vector at level $rs/2$. 

The relation between singular vectors and symmetric polynomials goes through the free-field representation. 
In the R sector, it is described in terms of the free-field modes $a_n$ and $b_n$, with $n\in\mathbb{Z}$, together with the vacuum charge operator $\pi_0$:
\begin{equation} \label{Ralgfock}
[a_n,a_m]=n\delta_{n+m,0} \, ,\qquad [a_0,\pi_0]=1\, ,\qquad \{b_n,b_m\}= \delta_{n+m,0}\,.
\end{equation}
  We define a one-parameter family of highest-weight states as  $|\eta\R\equiv e^{\eta \pi_0}|0\R$ satisfying 
\begin{equation} a_0|\eta\R=\eta |\eta \R,\qquad a_n|\eta\R=0 \quad\text{and}\quad b_n|\eta
\R=0,\qquad\forall\;n>0. 
\end{equation}
This allows us to introduce  the Fock space $\mathscr{F}$ with highest weight $|\eta\R$ over the superalgebra \eqref{Ralgfock}: it is generated by all states off the form
\begin{equation}b_{ -\La_1 }\cdots b_{ -\La_m }a_{-\La_{m+1}}\cdots a_{-\La_\ell} |{\eta}\R
\end{equation}
where the labeling of the states  satisfies \eqref{eqspart}.  The following expressions yield a representation of the R sector on  $\mathscr{F}$:  
\begin{align} \label{ffrR} L_n&=-\gamma(n+1)a_n+\frac{1}{2}\sum_{k\in\mathbb{Z}}:a_ka_{n-k}:+ \frac{1}{4}\sum_{k\in\mathbb{Z}}\big(n-2k+\frac{1}{2}\big):b_kb_{n-k}:\\
 G_n&=-2\gamma\big(n+\frac{1}{2}\big)b_n+\sum_{k\in\mathbb{Z}}a_kb_{n-k} ,
\end{align}where  $\gamma$  is related to the central charge via $c=\tfrac{3}{2}-12 \gamma^2$.

We are now in position to formulate the correspondence between the  free-field modes  and the differential operators acting on the space of symmetric superpolynomials (cf. Section \ref{A2}), which space will be denoted by $\mathscr{R}$ . It reads: 
\begin{equation} \label{corR}    a_n\longleftrightarrow \begin{cases} \frac{(-1)^{n-1}}{\sqrt{\alpha}}p_{-n} & n<0 \\ \eta & n=0\\ 
{(-1)^{n-1}} n\sqrt{\alpha}\partial_n &n>0\end{cases}\qquad  b_n\longleftrightarrow \begin{cases} \frac{(-1)^n}{\sqrt{2}}\tilde p_{-n} & n<0 \\ \frac{1}{\sqrt{2}}(\tilde p_0 +\tilde \partial_0) & n=0\\  (-1)^{n}\sqrt{2}\tilde \partial_n &n>0\end{cases}\, .
\end{equation}
Two points are noteworthy: the absence of $\alpha$ factors in the representation of the $b$ modes and  the presence of the zero mode $b_0$ which is represented by a combination of {the fermionic polynomial $\ti p_0$} and its derivative.  Eq \eqref{corR} together with the identification  $|\eta\R\leftrightarrow 1$ induce  the following correspondence between $\mathscr{F}$ and $\mathscr{R}$:  
\begin{equation} \label{corR2} b_{-\La_1}\cdots b_{-\La_m}a_{-\La_{m+1}}\cdots a_{\La_\ell}|\eta\R\longleftrightarrow \zeta_\La p_\La\,
\qquad  \, ,\end{equation}
where 
\begin{equation}  \zeta_\La=\frac{(-1)^{|\La|-(\ell-m)}}{2^{m/2}\alpha^{(\ell-m)/2}}.\end{equation}
We stress that the level in the Fock space is  equal to the bosonic degree of the superpolynomials (namely, the sum of all parts of $\La$). 

The free-field representation \eqref{ffrR}  and the correspondence \eqref{corR} yield a representation  of the super-Virasoro generators in  the R sector as differential operators acting on the space $\mathscr{R}$ of symmetric superpolynomials. The two relevant expressions for our present purpose are
\begin{align}  \label{GLpol}
	 \mathcal{G}_1&=  {\sqrt{2}}(3\gamma-\eta) \tilde \partial_1+\sqrt{\frac{\alpha}{2}}(\tilde p_0+\tilde \partial_0)\partial_1+\sqrt{\frac{\alpha}{2}} \sum_{n\geq 1} n \tilde p_n\partial_{n+1}+\sqrt{\frac{2}{\alpha}}\sum_{n\geq 1}   p_n \tilde \partial_{n+1}\nonumber\\
	 \mathcal{L}_1&=  {\sqrt{\alpha}}(\eta-2\gamma)   \partial_1- {\frac{1}{2}}(\tilde p_0+\tilde \partial_0)\tilde \partial_1-\sum_{n\geq 1} n  p_n\partial_{n+1}-{\frac{1}{2}}\sum_{n\geq 1}  (2n+1) \tilde p_n  \tilde \partial_{n+1}.
	\end{align}
We end up with the following correspondence between states of the module $ \mathscr{M}$ with highest-weight  state  $|h\R$ and symmetric superpolynomials: 
\begin{align}   \label{corrhwR} 
 \sum_\La c_\La G_{-\La^a}L_{-\La^s}|h\R& \longleftrightarrow \sum_\La c_\La \mathcal{G}_{-\La^a}\mathcal{L}_{-\La^s}(1)
\end{align}
where 
\begin{equation} h=\frac{1}{2}{\eta}(\eta-2\gamma)+\frac{1}{16}. \end{equation}

\subsection{Singular vector as superpolynomials}

In order to apply the correspondence \eqref{corrhwR}  to the $(r,s)$-{type} Kac module, we must  set \cite{DLM_jhep}
\begin{equation} \label{parametsing}
t=\alpha,\qquad\gamma=\frac{1}{2\sqrt{\alpha}}(\alpha-1) \qquad \text{and}\qquad \eta\,
{\equiv \eta_{r,s}}=\frac{1}{2\sqrt{\alpha}}\left((r+1)\alpha-(s+1)\right) \,.\end{equation}

In what follows,  $P_\La=P^{(\alpha)}_\La$ denotes  the sJack with parameter $\alpha$ and indexed by the superpartition $\La$, while $v_\Lambda=v_\La^{(\alpha)}$ stands for some coefficient depending rationally on $\alpha$.  
With the above parametrization, the positive-chirality R singular vector $\ket{\chi_{r,s}}$ can be represented as a superpolynomial 
\begin{equation} 
\label{defv}F_{r,s}= \sum_{\substack{\Lambda\\ m=0\,\text{mod}\,2\\|\La|=rs/2}}v_\Lambda  P^{}_\La,
\end{equation}
 if and only if 
 \begin{equation}  {\mathcal{G}_1(F_{r,s})=0\qquad \text{and}\qquad  \mathcal{L}_1(F_{r,s})=0 \, . }
\label{2sv}\end{equation}
The objective is to first characterize more precisely those $\La$ for which $v_{\La}\ne 0$ and then to find the explicit expression of $v_{\La}$. The formula to be presented below applies to the case $r$ odd. Recall that in the R sector, $r+s$ is odd. Therefore, in other to cover all situations, we need to be able to recover $F_{s,r}$ from $F_{r,s}$. This point is discussed in the following section.


\section{General formula for the coefficients of the sJacks-form of the Ramond singular vectors}

In this section, we present a conjecture for the explicit form of the coefficient $v_\Lambda$ in the representation of 
\begin{align}
	\ket{\chi_{r,s}}\longleftrightarrow \sum_{\Lambda \in \mathcal{A}_{r,s}}v_\Lambda P^{}_\Lambda.
\end{align}
 $\mathcal{A}_{r,s}$ is the set of all allowed superpartitions  of type $(r,s)$ defined below.

For convenience, we will work with singular vectors for which $r$ is odd and $s$ even. The case $r$ even and $s$ odd is recovered by a duality transformation that  exchanges $r\leftrightarrow s$,  as detailed in  Section \ref{sectionduality}.

\subsection{Allowed superpartitions}

As  previously observed in \cite{DLM_jhep}, those $\La$ that contribute to the  R singular vector at level $rs/2$ belong to  the set of $(r,s)$-self-complementary superpartitions of the R sector, which we denote by $\AR$  (cf.  Section B.4 in \cite{DLM_jhep}).  A superpartition $\Lambda$ belongs to  $\AR$ if  and only if    the complement of the diagram
$\Lambda^*$, in the rectangle\footnote{To be clear, the leftmost upper corner of the rectangle is adjusted with the external boundaries of the $(1,1)$ box of $\La^*$.} with $r$ columns and $s$ rows, corresponds to
$\Lambda^*$ rotated by 180  degrees.  Note that 
\begin{equation} \Lambda\in\AR \quad  \Longrightarrow  \quad |\Lambda|= \frac{rs}{2} .
\end{equation}

Consider for instance the case with $(r,s)=(3,4)$.  Suppose moreover that the fermionic degree is even.  Then, there are 40 superpartitions of appropriate degree.  However,   as illustrated below,   only 14 superpartitions amongst them   are  $(3,4)$-self-complementary: \footnote{This example was considered in Section B.4 of \cite{DLM_jhep} but the formulation of the rule for allowed superpartitions misses a condition that was tacitly assumed in the example (B.25) (namely, the third condition in eq. \eqref{selec} below). We thus rework this example properly.}
\begin{equation}
\tableau[scY]{&&\\&&\\\tf&\tf&\tf\\\tf&\tf&\tf\\}
\longleftrightarrow\;
\footnotesize{ \begin{matrix}(3,3)\\(3,0;3)\end{matrix}}\qquad
\normalsize{\tableau[scY]{&&\\&&\tf\\&\tf&\tf\\\tf&\tf&\tf\\}
\longleftrightarrow}\;
\footnotesize{ \begin{matrix}(3,2,1)\\(1,0;3,2)\\(2,0;3,1)\\(2,1;3)\\(3,0;2,1)\\(3,1;2)\\(3,2;1)\\(3,2,1,0;)\end{matrix} } \qquad \normalsize{
\tableau[scY]{&&\tf\\&&\tf\\&\tf &\tf\\&\tf&\tf\\}
\longleftrightarrow }\;
\footnotesize{ \begin{matrix}(2,2,1,1)\\(2,1;2,1)\\(2,0;2,1,1)\\ (1,0;2,2,1)\end{matrix}}\qquad
\end{equation}
where the boxes marked with thick frames correspond to the boxes of the rotated copy of $\La^*$.

The set  of allowed superpartitions of type $(r,s)$ for $r$ odd, denoted $\mathcal{A}_{r,s}$, is the set of all superpartition $\Lambda$ satisfying the following selection rules: 
\begin{subequations}\label{selec}
	\begin{align}
			&m=0\, \mathrm{mod} \, 2,  \\
		&\Lambda \in  \AR, \\
		&\ell(\Lambda) \leq s.
	\end{align}
\end{subequations}
In other words, $\La$ is an allowed superpartition of type $(r,s)$ if and only if $\La$ contains an even number of circles, the complement of $\La^*$ in the rectangle of width $r$ and height $s$ is a copy of $\La^*$,  and the only cell of $\La$ that can be out of the latter region is a circle in the first row.\footnote{For $r$ even, the only cell of $\La$ that can be out of the $r\times s$ rectangle is a circle in the first column.} 
This, as we said, defines the set  $\mathcal{A}_{r,s}$:
\beq
\text{$r$ odd}:\quad \mathcal{A}_{r,s}=\{\La\,|\,\Lambda \in  \AR,\, \text{$m$ even},\,\ell(\Lambda) \leq s\}.\eeq

Returning to the example for which $(r,s)=(3,4)$ we see that the superpartitions $(2,0;2,1,1)$ and $(1,0;2,2,1)$ are not allowed since they contradict the third selection rule; there are thus 12 allowed superpartitions of type $(3,4)$.

\subsection{The recursive structure of the allowed diagrams}

Since the allowed superpartitions must be self-complementary -- disregarding
the circles -- and  fit within the $r\times s$ rectangle, we can view any  diagram at a given level as being build up from a lower level one, by the addition of either a column or a row. For instance, $(r,s)$-type diagrams 
 are allowed to be wider than the $(r-2,s)$ ones. But the self-complementary requirement constrains their horizontal extension: every $(r-2,s)$ diagram can be transformed into a $(r,s)$ diagram by simply adding a column of length $s$ on the left side of the diagrams.  However, not all the allowed $(r,s)$-level diagrams are generated by a column adjunction. Another generating source comes from the vertical extension of the $(r,s-2)$-level ones. The ``rectangle-fitting" rule and  self-complementarity imply that this vertical extension must be done in a very specific way, namely by adding a row of length $r$ atop the diagram. \\

Let the core of $ \mathcal{A}_{r,s}$, denoted $ \mathcal{A}^*_{r,s}$,
 be the set of $\La\in  \mathcal{A}_{r,s}$ such that $\La=\La^*$ (i.e., those diagrams without circles).
Summarizing what has been said so far, the whole set $ \mathcal{A}^*_{r,s}$ can be obtained by adding a column of length $s$ to every elements of $ \mathcal{A}^*_{r-2,s}$ and adding a row of length $r$ to those of $ \mathcal{A}^*_{r,s-2}$.  The full set
 $ \mathcal{A}_{r,s}$ is recovered by adding pairs of circles to elements of $ \mathcal{A}^*_{r,s}$ in all the allowed way (including no circle addition).\\

A simple illustration of this recursive process is presented in 
the following example:
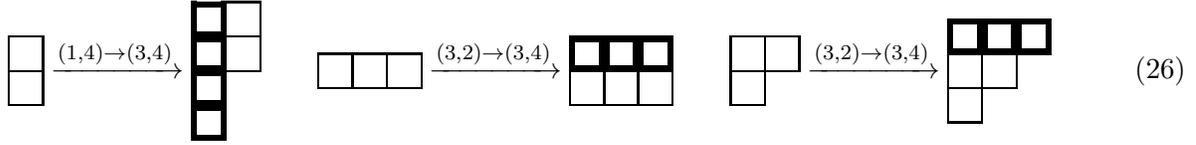
\beq\label{ex123}
\tableau[scY]{\\ \\}\xrw{(1,4)\rw(3,4)} \tableau[scY]{\tf&\\\tf&\\\tf\\\tf}\qquad 
\tableau[scY]{&&\\}\xrw{(3,2)\rw(3,4)}\tableau[scY]{\tf&\tf&\tf\\ &&}\qquad
\tableau[scY]{&\\ \\}\xrw{(3,2)\rw(3,4)} \tableau[scY]{\tf&\tf&\tf\\ &\\ \\}\qquad
\eeq
In this example, we obtain $ \mathcal{A}^*_{3,4}$ by adding one column of length $s=4$ to the unique element of $ \mathcal{A}^*_{1,4}$ displayed at the left in \eqref{ex123}, and adding a row of width $r=3$ atop the two elements of $ \mathcal{A}^*_{3,2}$. The result is $ \mathcal{A}^*_{3,4}=\lbrace (2^2,1^2),(3^2),(3,2,1) \rbrace$. In order to  generate the full set $\mathcal{A}_{3,4}$, we decorate these core-diagrams with pairs of circles in all allowed ways, obtain thus
\begin{align}
	\mathcal{A}_{3,4} :
	\begin{array}{l}
		\left\lbrace (3,3),(2,2,1,1),(3,2,1),(3,0;3),(2,1;2,1),(1,0;3,2),(2,0;3,1),\right. \\
		\left. (3,0;2,1),(2,1;3),(3,1;2),(3,2;1),(3,2,1,0;) \right\rbrace.
	\end{array}
\end{align}

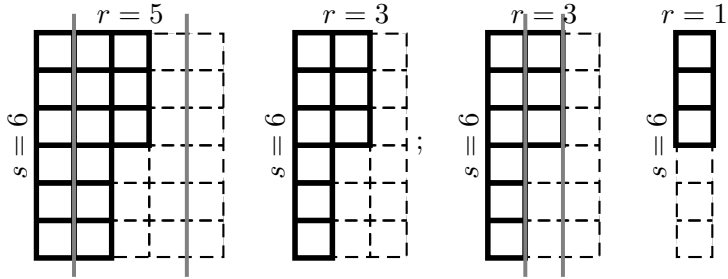
\begin{figure}[h]
\psscalebox{1}{
	\begin{pspicture}(0,-3)(10,1)
	\def\DashedSquare{\pspicture(.5,.5)\psframe[dimen=middle,linestyle=dashed](.5,.5)\endpspicture}
	\def\Square{\pspicture(.5,.5)\psframe[dimen=middle,linewidth=2pt](.5,.5)\endpspicture}
	\psboxfill{\DashedSquare}
	\psscalebox{1 -1}{\psframe[fillstyle=boxfill,linestyle=dashed](0,0)(2.5,3)}

	\psboxfill{\Square}
	\psscalebox{1 -1}{
	\pspolygon[fillstyle=boxfill,linewidth=2pt](0,0)(1.5,0)(1.5,1.5)(1,1.5)(1,3)(0,3)
	}
	\psline[linewidth=1.5pt,linecolor=gray](.5,.25)(.5,-3.25)
	\psline[linewidth=1.5pt,linecolor=gray](2,.25)(2,-3.25)
	
	\rput(3.5,0){
	\psboxfill{\DashedSquare}
	\psscalebox{1 -1}{\psframe[fillstyle=boxfill,linestyle=dashed](0,0)(1.5,3)}

	\psboxfill{\Square}
	\psscalebox{1 -1}{
	\pspolygon[fillstyle=boxfill,linewidth=2pt](0,0)(1,0)(1,1.5)(.5,1.5)(.5,3)(0,3)
	}
	}
	
	\rput(6,0){
	\psboxfill{\DashedSquare}
	\psscalebox{1 -1}{\psframe[fillstyle=boxfill,linestyle=dashed](0,0)(1.5,3)}

	\psboxfill{\Square}
	\psscalebox{1 -1}{
	\pspolygon[fillstyle=boxfill,linewidth=2pt](0,0)(1,0)(1,1.5)(.5,1.5)(.5,3)(0,3)
	}
	\psline[linewidth=1.5pt,linecolor=gray](.5,.25)(.5,-3.25)
	\psline[linewidth=1.5pt,linecolor=gray](1,.25)(1,-3.25)
	}
	
	\rput(8.5,0){
	\psboxfill{\DashedSquare}
	\psscalebox{1 -1}{\psframe[fillstyle=boxfill,linestyle=dashed](0,0)(.5,3)}

	\psboxfill{\Square}
	\psscalebox{1 -1}{
	\pspolygon[fillstyle=boxfill,linewidth=2pt](0,0)(.5,0)(.5,1.5)(0,1.5)
	}}	
	
	\rput(1.25,.25){$r=5$}
	\rput{90}(-.25,-1.5){$s=6$}
	\rput(4.25,.25){$r=3$}
	\rput[b]{90}(3.3,-1.5){$s=6$}
	\rput(5.1,-1.5){;}
	
	\rput(6.75,.25){$r=3$}
	\rput{90}(5.75,-1.5){$s=6$}
	\rput(8.75,.25){$r=1$}
	\rput{90}(8.25,-1.5){$s=6$}
	\end{pspicture}
	}

\caption{Illustration of the column-removal operation for self-complementary diagrams. Here, two removals of a column of length $6$ are displayed and after each operation, there results a self-complementary diagram.}\label{fig.r5s6tor1s6}
\end{figure}

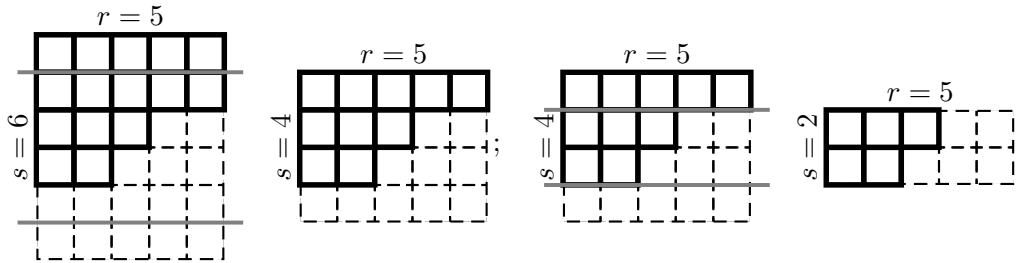
\begin{figure}[h]
\psscalebox{1}{
	\begin{pspicture}(0,-3)(13,1)
	
	\def\Square{\pspicture(.5,.5)\psframe[dimen=middle,linewidth=2pt](.5,.5)\endpspicture}
	\def\DashedSquare{\pspicture(.5,.5)\psframe[dimen=middle,linestyle=dashed](.5,.5)\endpspicture}
	
	\psboxfill{\DashedSquare}
	\psscalebox{1 -1}{\psframe[fillstyle=boxfill,linestyle=dashed](0,0)(2.5,3)}

	\psboxfill{\Square}
	\psscalebox{1 -1}{
	\pspolygon[fillstyle=boxfill,linewidth=2pt](0,0)(2.5,0)(2.5,1)(1.5,1)(1.5,1.5)(1,1.5)(1,2)(0,2)
	}
	\psboxfill{\DashedSquare}
	\rput(3.5,-.5){
	\psscalebox{1 -1}{\psframe[fillstyle=boxfill,linestyle=dashed](0,0)(2.5,2)}
	\psboxfill{\Square}
	\psscalebox{1 -1}{
	\pspolygon[fillstyle=boxfill,linewidth=2pt](0,0)(2.5,0)(2.5,.5)(1.5,.5)(1.5,1)(1,1)(1,1.5)(0,1.5)
	}
	}
	\psboxfill{\DashedSquare}
	\rput(7,-.5){
	\psscalebox{1 -1}{\psframe[fillstyle=boxfill,linestyle=dashed](0,0)(2.5,2)}
	\psboxfill{\Square}
	\psscalebox{1 -1}{
	\pspolygon[fillstyle=boxfill,linewidth=2pt](0,0)(2.5,0)(2.5,.5)(1.5,.5)(1.5,1)(1,1)(1,1.5)(0,1.5)
	}
	}
	\psboxfill{\DashedSquare}
	\rput(10.5,-1){
	\psscalebox{1 -1}{\psframe[fillstyle=boxfill,linestyle=dashed](0,0)(2.5,1)}
	\psboxfill{\Square}
	\psscalebox{1 -1}{
	\pspolygon[fillstyle=boxfill,linewidth=2pt](0,0)(1.5,0)(1.5,.5)(1,.5)(1,1)(0,1)
	}
	}
	\psline[linecolor=gray,linewidth=1.5pt](-.25,-.5)(2.75,-.5)
	\psline[linecolor=gray,linewidth=1.5pt](-.25,-2.5)(2.75,-2.5)
	\psline[linecolor=gray,linewidth=1.5pt](6.75,-1)(9.75,-1)
	\psline[linecolor=gray,linewidth=1.5pt](6.75,-2)(9.75,-2)	
	\rput(11.75,-.75){$r=5$}
	\rput{90}(10.25,-1.5){$s=2$}
	\rput(6.125,-1.5){;}
	\rput(8.25,-.25){$r=5$}
	\rput{90}(6.75,-1.5){$s=4$}
	\rput(4.75,-.25){$r=5$}
	\rput{90}(3.25,-1.5){$s=4$}
	\rput(1.25,.25){$r=5$}
	\rput{90}(-.25,-1.5){$s=6$}
	\end{pspicture}
	}
\caption{Illustration similar to Figure \ref{fig.r5s6tor1s6} but here for row removal.}\label{fig.r5s6tor5s2}
\end{figure}

{
The completeness of this recursive construction of the $\mathcal{A}^*_{r,s}$ set is based on the simple observation that a self-complementary diagram necessarily has a column of length $s$ or a row of length $r$ (but not both). This, in turn, implies that any self-complementary diagram can be deconstructed uniquely down to a simple self-complementary diagram --a single row diagram corresponding to $(r,s)=(2k+1,2)$ for some $k\geq 0$ --, by successively deleting a column (of length $s$) or a row (of length $r$) -- with the understanding that one of the values of $r$ and $s$ is changing at every step.
This is illustrated in Figures
\ref{fig.r5s6tor1s6} and \ref{fig.r5s6tor5s2}. The first of these figures shows the transformation of the core partition $(3^3,2^3)$ for $(r,s)=(5,6)$ down to the partition $(1^3)$ for $(r,s)=(1,6)$ by removing twice a column of length 6. Observe that at each step, the resulting core diagram is self-complementary. This is easily seen to be a consequence of the deconstruction mechanism. Indeed, the figure also displays the defining $r\times s$ rectangle in dashed lines. Now, in the column reduction process, two columns are removed from this rectangle, one at each extremity (at the exterior of the gray lines). But that shows clearly  that the partition and its complement in the $r\times s$ rectangle are reduced in exactly the same way, ensuring  the preservation of self-complementarity. That two columns are removed from the defining rectangle is the reason for which $r$ decreases by 2 at each step. Notice that the last resulting diagram can be further reduced by removing twice a row of length 1, ending up with the partition $(1)$ for $(r,s)=(1,2)$.
Figure  \ref{fig.r5s6tor5s2} illustrates the similar process of removing rows of length $r$. Notice again that the last diagram can be further reduced to a single box  by two column-removing steps.
It is clear that the choice of either a column or a row removal at a given step is determined by the partition under consideration and it is unambiguous. For instance, with a staircase diagram, these operations alternate.} \\

This recursive pattern allows us to identify every box of a diagram as belonging to a row or a column which has been inserted at a certain $(r,s)$ level. As we just argued, the increment in $r$ results from the addition of columns, while that  of $s$ is due to the addition of rows. Therefore, within an allowed $(r,s)$ diagram, we write $(\tr_j,\ts_i)$ in the box with coordinate $(i,j)$, where
\begin{align}
	\tilde{r}_{j} &\equiv r-2(j-1)\, , \\
	\tilde{s}_{i} &\equiv s-2(i-1) \, .
\end{align}
From these data, we now introduce two sets:
\begin{align}
	\mathcal{S}_{\Lambda,s}&=\{(i,j)\in\La\,|\, l_{\Lambda^\ast}(i,j)+1=\tilde{s}_{i}\}\, ,\nonumber\\
	\mathcal{R}_{\Lambda,r}&=\{(i,j)\in\La\,|\,a_{\Lambda^\ast}(i,j)+1=\tilde{r}_{j}\}\, .\end{align} 
It will also be convenient to denote by $\mathcal{O}_\Lambda$  the set of indices $(i,j)$ of every circle in the diagram $\Lambda$, or equivalently, the boxes of $\La^\cd$ that are not in $\La^*$:  
\begin{align} 
	\mathcal{O}_{\Lambda}&=\{(i,j)\in\La^\cd/\La^*\}.\end{align}
	
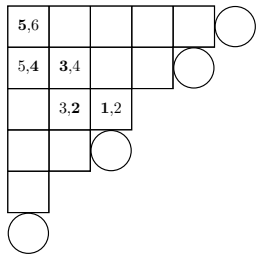
\begin{figure}[!h]
	\centering
	\psscalebox{.55}{
	\begin{pspicture}(0,0)(5,-6)
\def\Square{\pspicture(1,1)\psframe[dimen=middle](1,1)\endpspicture}
\psboxfill{\Square}
\psscalebox{1 -1}{%
\rput(.5,.5){\pspolygon[fillstyle=boxfill](0,0)(5,0)(5,1)(4,1)(4,2)(3,2)(3,3)(2,3)(2,4)(1,4)(1,5)(0,5)}
\pscircle(6,1){.5}
\pscircle(5,2){.5}
\pscircle(3,4){.5}
\pscircle(1,6){.5}
}
\rput(1,-1){\textbf{5},6}
\rput(1,-2){5,\textbf{4}}
\rput(2,-2){\textbf{3},4}
\rput(2,-3){3,\textbf{2}}
\rput(3,-3){\textbf{1},2}
\end{pspicture}
}
\caption[Illustration of sets $\mathcal{S}_{\Lambda,s}$, $\mathcal{R}_{\Lambda,r}$ and $\mathcal{O}_{\Lambda}$]{Illustration of sets $\mathcal{S}_{\Lambda,s}$, $\mathcal{R}_{\Lambda,r}$ and $\mathcal{O}_{\Lambda}$}
\label{fig.IllusEnsembles}
\end{figure}

Figure \ref{fig.IllusEnsembles} illustrates these definitions. It displays a diagram in which we have identified boxes with their $(\ti r_j, \ti s_i)$ coordinates. For those  boxes that belong to the set $\mathcal{R}_{\Lambda,r}$, the $\ti r$ coordinate is bold, while for those which are elements of the set $\mathcal{S}_{\Lambda,s}$, the $\ti s$ coordinate is bold. We then see that $\mathcal{R}_{(5,4,2,0;3,1),5}=\lbrace (1,1),(2,2),(3,3) \rbrace$, $\mathcal{S}_{(5,4,2,0;3,1),6}=\lbrace (2,1), (3,2) \rbrace$. Finally, the set $\mathcal{O}_{(5,4,2,0;3,1)}$ is clearly $\lbrace (1,6), (2,5), (4,3), (6,1) \rbrace$.
\subsection{The explicit formula for the singular-vector expansion coefficients}

We are now in position to present the explicit conjectural expression for the coefficients $v_\La$ in \eqref{defv}:
\begin{align}
	v_{\Lambda}(r,s) &= \epsilon(\Lambda,r,s)B(\Lambda,r,s)C(\Lambda,r,s),
\end{align}
where  $B(\Lambda,r,s)$ is a function that captures the part of the coefficient that depends solely on box structure of the diagram, while  $C(\Lambda,r,s)$ depends upon the diagram's circle pattern. 
Finally, the factor $\epsilon(\Lambda,r,s)$, which is explained latter in Section \ref{SectSign},  is a sign that also depends upon the circles arrangement.\\

In the following expressions, in order to shorten the notation, we denote a box by $t$ when the specification of its coordinates are not necessary. For $t=(i,j)$, $\tr_t=\tr_j$ and $\ts_t=\ts_i$.
The coefficient $B$ reads
\begin{align}
	B(\Lambda,r,s)&= \prod_{t \in \mathcal{S}_{\Lambda,s}} k(t)f(\tilde{r}_t,\tilde{s}_t)\, , \end{align}
	where
	\begin{align}
	&k(t) = \prod_{n=0}^{l_{\Lambda^\ast}(t)} 
	{h^{\downarrow\uparrow}_\Lambda(i+n,j)}\, , \\ 
	&{h^{\downarrow\uparrow}_\Lambda(t)}=\frac{h^\downarrow_\Lambda(t)}{h^\uparrow_\Lambda(t)} \, , \\
	\label{hdn}
&h^\downarrow_\La(t)=l_{\La^*}(t)+1+\a\,a_{\La^\cd}(t) \, , \\
\label{hup}
&h^\uparrow_\La(t)=l_{\La^\cd}(t)+\a(a_{\La^*}(t)+1)\, , 
\end{align}
and
\begin{align}
	f(r,s)&=\frac{(r-1)\alpha}{(r-1)\alpha + s}\prod_{\substack{i=s-1 \\ i \text{ odd}}}^1 \frac{r\alpha + i}{(r-2)\alpha + i}\, .
\end{align}
The $C(\La,r,s)$ coefficient is 
\begin{align}
	C(\Lambda,r,s)& = \prod_{{t=}(i,j) \in \mathcal{R}_{\Lambda,r}} \left\lbrace 
	 \prod_{\substack{(i^\prime,j^\prime) \in \mathcal{O}_\Lambda\\ 0 < j-j^\prime+ \tilde{r}_{j} \leq \tilde{r}_{j}}} \frac{||(i,j+\tilde{r}_{j}),(i^\prime-1,j^\prime)||^{(1-{\delta_{\ti l(t)+1,\tilde{s}_{i}})}}}
	 {||(i,j+\tilde{r}_{j}),(i^\prime,j^\prime)||^{(1-{\delta_{\ti a(t),\tilde{r}_{j}})}}} \right. \nonumber \\
	&\quad\times \left.
	\prod_{\substack{(i^\prime,j^\prime) \in \mathcal{O}_\Lambda \\ 0 \leq j^\prime - j < \tilde{r}_{j}}} \frac{||(i-1+\tilde{s}_{i},j),(i^\prime-1,j^\prime)||^{(1-{\delta_{\ti a(t),\tilde{r}_{j}})}}}
	{||(i-1+\tilde{s}_{i},j),(i^\prime,j^\prime)||^{(1-{\delta_{\ti l(t)+1,\tilde{s}_{i}})}}} \right\rbrace,
\end{align}
where $\ti a=a_{\La^\cd}$ and $\ti l=l_{\La^\cd}$.
Here, $ \|t_1,t_2\|$ refers  to the $\alpha$-distance between
$t_1$ and $t_2$: for $t_1=(i_1,j_1)$ and $t_2=(i_2,j_2)$,
this distance is defined as
\beq \|t_1,t_2\|=| i_1-i_2 + \alpha(j_2-j_1)|.
\eeq

\subsection{Illustration of the combinatorics underlying the general formula}

Since the explicit formula has a complicated looking form, we will illustrate it by considering separately its three components, the two functions $B(\Lambda,r,s)$ and $C(\Lambda,r,s)$, and finally, the sign $\epsilon(\La,r,s)$.  

\subsubsection{The $B(\Lambda,r,s)$ function}
The $B(\La,r,s)$ function is built up from the elements of the set $\mathcal {S}(\La,s)$. 
For each box $t=(i,j)$ such that $\tilde{s}_{i}-1$ is equal to the number of boxes below it, there is a contributing factor $f(\tilde{r}_{j},\tilde{s}_{i})$ multiplied by the function $k(\Lambda,i,j)$. The latter is given by the product of ${h^\downarrow}(t')/{h^\uparrow}(t')$ for $t'=t$ and  every box below $t$: 
\beq t'=\{(i',j)\, |\, i\leq i'\leq i+l_{\La^*}(t)\}\eeq
If the set  $\mathcal {S}(\La,s)$ is empty, it is understood that $B(\Lambda,r,s)=1$.\\

\begin{figure}[!h]
\centering
	\psscalebox{.7}{
	\begin{pspicture}(0,0)(7,12)
\rput(-0.5,.5){\psframe[linestyle=none,hatchcolor=lightgray,fillstyle=vlines,fillcolor=lightgray](0,11)(1,1)}
\rput(-0.5,.5){\psframe[linestyle=none,hatchcolor=lightgray,fillstyle=vlines,fillcolor=lightgray](1,9)(2,3)}
\rput(-0.5,.5){
\psline(0,12)(7,12)
\psline(0,11)(7,11)
\psline(0,10)(6,10)
\psline(0,9)(6,9)
\psline(0,8)(5,8)
\psline(0,7)(5,7)
\psline(0,6)(5,6)
\psline(0,5)(2,5)
\psline(0,4)(2,4)
\psline(0,3)(2,3)
\psline(0,2)(1,2)
\psline(0,1)(1,1)

\psline(0,12)(0,1)
\psline(1,12)(1,1)
\psline(2,12)(2,3)
\psline(3,12)(3,6)
\psline(4,12)(4,6)
\psline(5,12)(5,6)
\psline(6,12)(6,9)
\psline(7,12)(7,11)
}
\pscircle(7,12){.5}
\pscircle(1,3){.5}
\rput(0,12){7,12}
\rput(0,11){\underline{7,\textbf{10}}}
\rput(1,11){5,10}
\rput(1,10){5,8}
\rput(1,9){\underline{5,\textbf{6}}}
\rput(2,9){3,6}
\rput(2,8){3,4}
\rput(2,7){3,2}
\end{pspicture}
}
\caption[Illustration of the $B(\Lambda,r,s)$ coefficient]{}
	\label{IllustrationB}
\end{figure}

Consider the superpartition $\La=(7,1;6,6,5,5,5,2,2,2,1)$ whose diagram is displayed in Figure \ref{IllustrationB}.
This is easily checked to be an element of $\mathcal{A}_{7,12}$. 
We first label the boxes by their indices $(\tilde{r}_{j},\tilde{s}_{i})$. 
A partial label-filling is illustrated in the figure. Those boxes that are elements of $\mathcal {S}_{\La,12}$ have their labels underlined. So here 
$\mathcal {S}_{\La,12}=\lbrace (2,1),(4,2)\rbrace$. There are thus two contributing factors $f(\tilde{r}_{j},\tilde{s}_{i})$, namely $f(7,10)$ and $f(5,6)$. This 
is multiplied by  the ratios ${h^\downarrow(i,j)}/{h^\uparrow (i,j)}$ of those underlined-label boxes and each one underneath, which corresponds  to the hatched boxes in the figure. Multiplying all these factors yields $B(\La,7,12)$:
\begin{align}
	&B((7,1;6,6,5,5,5,2,2,2,1),7,12) \nonumber \\
	&=f(7,10){h^{\downarrow\uparrow}(2,1) h^{\downarrow\uparrow}(3,1)\dots h^{\downarrow\uparrow}(11,1)}
	\times f(5,6) {h^{\downarrow\uparrow}(4,2) h^{\downarrow\uparrow}(5,2) \dots h^{\downarrow\uparrow}(9,2)}
	\nonumber \\
	& = \frac { \left( 7\,\alpha+9 \right)  \left( \alpha+2 \right) ^{2} \left( 4\,\alpha +7\right)  \left( \alpha+5 \right)  \left( \alpha+4 \right) }{ \left( 3\,\alpha+1 \right)  \left( 3\,\alpha+5 \right)  \left( 5\,\alpha +7\right) ^{2}}  \times \frac { \left( 7\,\alpha+5 \right)  \left( 7\,\alpha+3 \right)  \left( 7\,\alpha+1 \right) \alpha}{ \left( 2\,\alpha+3 \right) ^{3} \left( 5\,\alpha+6 \right)  \left( \alpha+1 \right) ^{5} \left( 4\,\alpha+5 \right) }.	\label{eq.IllustrationBcoeff}
\end{align}


\subsubsection{The $C(\Lambda,r,s)$ function}
To calculate the part of the coefficient associated with $C(\Lambda,r,s)$, we first identify the boxes for which the number $\tilde{r}_{j}$ correspond to the arm of the box plus 1. 
Corresponding to each of those boxes, we add two triangles (one up, one down), each equipped with two labels: $\vartriangle_{\tilde{r}_{j},\tilde{s}_{i}}$ in position $(i,j+\tilde{r}_{j})$ and $\triangledown_{\tilde{r}_{j},\tilde{s}_{i}}$ in position $(i-1+\tilde{s}_{i},j)$. The function $C(\La,r,s)$ is built from the various (shifted) $\a$-distance between these triangles and the circles in the diagram of $\La$, as we now detail. \\

{
Note that by construction, there is a triangle (of either type) at the end of each row (but not necessarily at the end of each column) and every circle becomes filled with a triangle.
Let us first argue that there is a triangle at the end of each row of a core diagram. Since any diagram of $\mathcal{A}_{r,s}$ is obtained by adding circles to elements of $\mathcal{A}^*_{r,s}$, the second statement (that circles are filled by triangles) will follow. The argument relies on the recursive
structure of self-complementary diagrams. The starting point is a row diagram with $2k+1$ boxes, corresponding to $(r,s)=(2k+1,2)$. We add a triangle at position $(1,2k+2)$ and a reversed one at position $(2,1)$. Keeping adding rows amounts to add a triangle at the end of each new row and pile reversed triangles at positions $(4,1),\, (6,1),\cdots (s',1)$.  See for instance the first three steps in \eqref{extra} below, where the beginning of the recursive reconstruction of $\La^*=(7,6,6,5,5,5,2,2,2,1,1)\in\mathcal{A}^*_{7,12}$ is displayed. Adding a column of length $s'$ shift this pile of down triangles to the end of length-one rows, as in the last step in the following example:
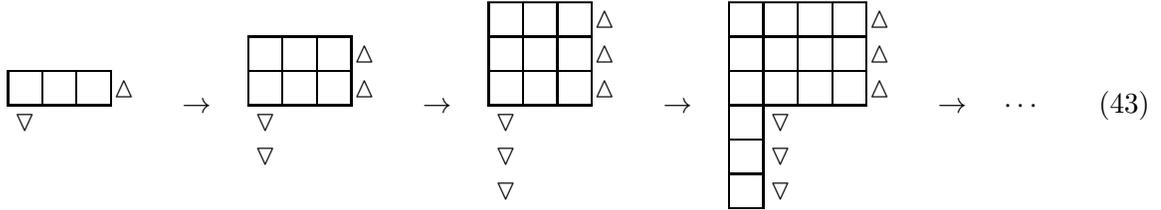
\beq\label{extra}
{\tableau[scY]{&&&\bl\vartriangle{} \\ \bl\triangledown}}\quad\rw\quad
{\tableau[scY]{&&&\bl\vartriangle{} \\&&&\bl\vartriangle{} \\ \bl\triangledown\\ \bl\triangledown}}
\quad\rw\quad
{\tableau[scY]{&&&\bl\vartriangle{} \\&&&\bl\vartriangle{} \\&&&\bl\vartriangle{} \\ \bl\triangledown\\ \bl\triangledown\\ \bl\triangledown}}
\quad\rw\quad
{\tableau[scY]{&&&&\bl\vartriangle{} \\&&&&\bl\vartriangle{} \\&&&&\bl\vartriangle{} \\& \bl\triangledown\\& \bl\triangledown\\ &\bl\triangledown}}
\quad\rw\quad\cdots
\eeq
Adding more columns of length $s'$ just push these triangles further right. Since the addition of rows and columns are the two basic operations for building up self-complementary core diagrams, this shows that each row of such a core diagram ends with a triangle, up or down.}\\

The numerator of each product is the `upper-shifted' $\a$-distance between every triangle and 
every circle. The upper-shifting means that the $\a$-distance is calculated not directly with a circle but rather with the box just above it (and the case where there  is no box above the circle is treated below).
The first numerator captures the contributions of the distances between $\vartriangle_{\tilde{r}_{j},\tilde{s}_{i}}$ and the circles.  
At first,  the exponent $(1-\delta_{\ti l(t)+1,\tilde{s}_{i}})$ indicates that
 if $\triangledown_{\tilde{r}_{j},\tilde {s}_{i}}$ is circled, this whole numerator factor reduces to 1.
Otherwise,
the constraint $0<j-j'+\ti r_j\leq \ti r_j$ means that only those circles that are southwest of $\vartriangle_{\tilde{r}_{j},\tilde{s}_{i}}$ and not horizontally further apart than $\ti r_j$ from  $\vartriangle_{\tilde{r}_{j},\tilde{s}_{i}}$ do contribute. 
The second numerator factor keeps track of the shifted $\a$-distances between the reversed triangles $\triangledown_{\tilde {r}_{j},\tilde {s}_{i}}$ and the circles. Again,  if $\vartriangle_{\tilde {r}_{j},\tilde {s}_{i}}$ is circled, this whole factor reduces to one. 
(This condition takes cares of the situation where there is no box above a circle, that is, the first row ends with a circle.) The condition $0\leq j'-j<\ti r_j$ states that only those circles above or northeast of $\triangledown_{\tilde {r}_{j},\tilde {s}_{i}}$ contribute, as long as their horizontal separation is strictly smaller than $\ti r_j$.
The denominator factors are similar except that the upper-shifted $\a$-distance is replaced by the unshifted $\a$-distance and if $\vartriangle$ or $\triangledown$ is circled, the corresponding factor is 1, i.e., the $\a$-distances from $\vartriangle$ or $\triangledown$ do not contribute).\\

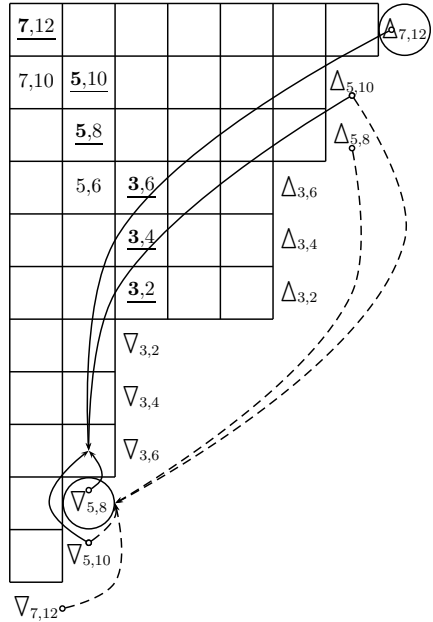
\begin{figure}[!h]
	\centering
	\psscalebox{.7}{
	\begin{pspicture}(0,0)(7,12)
\rput(-0.5,.5){
\psline(0,12)(7,12)
\psline(0,11)(7,11)
\psline(0,10)(6,10)
\psline(0,9)(6,9)
\psline(0,8)(5,8)
\psline(0,7)(5,7)
\psline(0,6)(5,6)
\psline(0,5)(2,5)
\psline(0,4)(2,4)
\psline(0,3)(2,3)
\psline(0,2)(1,2)
\psline(0,1)(1,1)

\psline(0,12)(0,1)
\psline(1,12)(1,1)
\psline(2,12)(2,3)
\psline(3,12)(3,6)
\psline(4,12)(4,6)
\psline(5,12)(5,6)
\psline(6,12)(6,9)
\psline(7,12)(7,11)
}
\pscircle(7,12){.5}
\pscircle(1,3){.5}
\rput(0,12){\underline{\textbf{7},12}}
\rput(0,11){7,10}
\rput(1,11){\underline{\textbf{5},10}}
\rput(1,10){\underline{\textbf{5},8}}
\rput(1,9){5,6}
\rput(2,9){\underline{\textbf{3},6}}
\rput(2,8){\underline{\textbf{3},4}}
\rput(2,7){\underline{\textbf{3},2}}
\rput(7,12){$\bigtriangleup_{7,12}$}\rput(0,1){$\bigtriangledown_{7,12}$} \pscurve[linestyle=dashed]{o->}(.5,1)(1.5,1.5)(1.5,3) \pscurve{o->}(6.75,12)(1.5,8)(1,4)
\rput(6,11){$\bigtriangleup_{5,10}$}\rput(1,2){$\bigtriangledown_{5,10}$} \pscurve[linestyle=dashed]{o->}(6,10.75)(7,8)(1.5,3) \pscurve[linestyle=dashed]{o->}(1,2.25)(1.5,2.75)(1.5,3) \pscurve{o->}(1,2.25)(.25,3)(1,4) \pscurve{o->}(6,10.75)(1.5,7)(1,4)
\rput(6,10){$\bigtriangleup_{5,8}$}\rput(1,3){$\bigtriangledown_{5,8}$} \pscurve[linestyle=dashed]{o->}(6,9.75)(6,7)(1.5,3) \pscurve[linestyle=solid]{o->}(1,3.25)(1.25,3.5)(1,4) 
\rput(5,9){$\bigtriangleup_{3,6}$}\rput(2,4){$\bigtriangledown_{3,6}$}
\rput(5,8){$\bigtriangleup_{3,4}$}\rput(2,5){$\bigtriangledown_{3,4}$}
\rput(5,7){$\bigtriangleup_{3,2}$}\rput(2,6){$\bigtriangledown_{3,2}$}
\end{pspicture}
	}
	\caption[Illustration of the $C(\Lambda,r,s)$ coefficient]{Dressing of the diagram of Figure \ref{IllustrationB} with  the $\vartriangle,\triangledown$-terms, together with the indication of the factors contributing to the numerator (solid lines) and the denominator (dashed lines) of the $C(\La,r,s)$ coefficient.}\label{IllusC}
\end{figure}

We illustrate these rules by considering again the case $\La=(7,1;6,6,5,5,5,2,2,2,1)$ whose diagram, augmented with the $\vartriangle,\triangledown$-factors inserted, is presented in Figure \ref{IllusC}.
 The solid lines link the pairs ($\vartriangle$, box-above-a-circle) or ($\triangledown$, box-above-a-circle) that contribute to  the numerator and the dashed lines similarly related pairs ($\vartriangle$, circle) or ($\triangledown$, circle) that contribute to the denominator. 
 Consider first  $\vartriangle_{7,12}$: there is one solid line linking it to the box atop the circle $(10,2)$; its contribution is $6\a+8$. There is no dashed line originating from  $\vartriangle_{7,12}$  because it is circled. From  $\vartriangle_{5,10}$ originates  one line of each type for a factor $(5\a+7)/(5\a+8)$. 
 The circle in position $(10,2)$ is separated from $\vartriangle_{3,n}$ for $n=2,4,6$ by an horizontal distance 4; since this is larger than $\ti r_j=3$, these terms do not contribute. A similar conclusion holds for the three terms $\triangledown_{3,n}$. A solid line starts from $\triangledown_{5,8}$ and ends on the box just above, for a contribution of  1. 
A line of each type originates 
from  $\triangledown_{5,10}$ for a factor $2/1$. 
Finally, for $\triangledown_{7,12}$ there is no numerator term (no solid line) $\vartriangle_{7,12}$ is circled. The dashed line that starts from there only goes to the $(10,2)$ circle since the $(1,8)$ circle is at an horizontal distance of $\tilde {r}_{j}=7$. This yields $1/(\a+2)$.
Collecting these factors yields:
\beq C((7,1;6,6,5,5,5,2,2,2,1),7,12)=\frac{2(6\alpha+8)(5\alpha+7)}{(5\alpha+8)(5\alpha+7)(\alpha+2)}.
\eeq


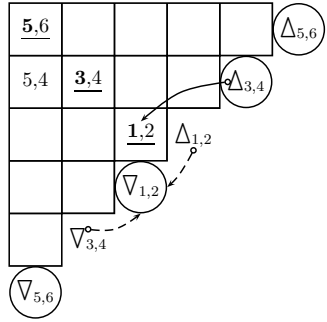
\begin{figure}[!h]
	\centering
	\psscalebox{.7}{
	\begin{pspicture}(0,0)(5,-6)
\def\Square{\pspicture(1,1)\psframe[dimen=middle](1,1)\endpspicture}
\psboxfill{\Square}
\psscalebox{1 -1}{%
\rput(.5,.5){\pspolygon[fillstyle=boxfill](0,0)(5,0)(5,1)(4,1)(4,2)(3,2)(3,3)(2,3)(2,4)(1,4)(1,5)(0,5)}
\pscircle(6,1){.5}
\pscircle(5,2){.5}
\pscircle(3,4){.5}
\pscircle(1,6){.5}
}
\rput(1,-1){\underline{\textbf{5},6}}
\rput(1,-2){5,4}
\rput(2,-2){\underline{\textbf{3},4}}
\rput(3,-3){\underline{\textbf{1},2}}
\rput(6,-1){$\bigtriangleup_{5,6}$}\rput(1,-6){$\bigtriangledown_{5,6}$}
\rput(5,-2){$\bigtriangleup_{3,4}$}\rput(2,-5){$\bigtriangledown_{3,4}$}
\rput(4,-3){$\bigtriangleup_{1,2}$}\rput(3,-4){$\bigtriangledown_{1,2}$}
\pscurve[linestyle=dashed]{o->}(4,-3.3)(3.7,-3.8)(3.5,-4)
\pscurve[linestyle=dashed]{o->}(2,-4.8)(2.5,-4.8)(3,-4.5)
\pscurve[linestyle=solid]{o->}(4.65,-2)(3.75,-2.2)(3,-2.75)
\end{pspicture}
}
\caption[Illustration of the $C(\Lambda,r,s)$ coefficient for a 4 circle diagram]{Another illustration of the triangles insertion and the arrow-contribution indicators, this time for a diagram with four circles}
\label{fig.IllusC4cercles}
\end{figure}

It is clear from the above rules that for a given core diagram ($\La^*$), the more circles there are in $\La$, the least  number of factors contribute to $C(\La,r,s)$. We illustrate this in Figure \ref{fig.IllusC4cercles} which displays a diagram with six rows but four of them ending with a circle. Let us detail the contributing terms. Both the $(5,6)$-indexed  triangles are circled, so they do not contribute. $\vartriangle_{3,4}$ being  circled  cancels the solid lines of $\triangledown_{3,4}$ and its own dashed lines. Only the solid line linking it to the box in position $(3,3)$ contributes and this for a factor $(2 \alpha +1)$. Finally we only need to consider the two  dashed lines ending on $\triangledown_{1,2}$; these contribute to $1/{(\alpha+1)^2}$. The corresponding $C(\La,r,s)$ factor  for this diagram is thus:
\begin{align}
C((5,4,2,0;3,1),5,6)=\frac{(2 \alpha +1)}{(\alpha+1)^2}.
\end{align}

\subsubsection{The $\epsilon$ function} \label{SectSign}
As already indicated, each circle is filled by a triangle, either up or down. It turns out that the sign associated to a given diagram is fixed by the triangle-data in the circles. Let us associate a degree to a triangle, up or down, given by the product of its labels divided by  2 and indicate the degree as $|\!\vartriangle\!|$ and $|\triangledown|$. We now introduce two quantities:
\begin{align}
n_\vartriangle&=\text{number of circled $\vartriangle$},\nonumber\\
p_\diamond&=\text{number of circled pairs $(\vartriangle,\triangledown)$ 
 such that $ | {\vartriangle}_{\ti r,\ti s} | < | {\triangledown}_{\ti r^\prime, \ti s^\prime} |$}.
\end{align}
Then, the sign of a diagram with $m$ circles
 reads
 \beq \epsilon(\La,r,s)=(-1)^{\binom{m}{2}+\binom{n_\vartriangle}{2}+\, p_\diamond},\eeq
where it is understood that $\binom12=0$.  For the example of Figure \ref{IllusC}, $m=2,\,n_\vartriangle=1 $, and because $|\triangledown_{5,8}|<|\!\!\vartriangle_{7,12}\!\!|$, $p_\diamond=0$, so that $\epsilon=1$.  For the diagram in Figure \ref{fig.IllusC4cercles}, we have $m=4$, $n_\vartriangle=2 $ and $p_\diamond=1$, which yields $\epsilon=+1$.

\subsubsection{The formula for $\ket{\chi_{1,s}}$} 
As a simple application of this general formula, we can readily obtain the superpolynomial representation of $\ket{\chi_{1,s}}$. 
Given that $\mathcal{A}_{1,s}=\lbrace (1^{s/2}), (1,0;1^{s/2-1})\rbrace$, the set  $\mathcal{S}_{\Lambda,s}=\emptyset$, so that $B(\La,1,s)=1$. Similarly, it is a simple execise  to prove that the coefficient $C(\La,1,s)=1$.  Hence, only the sign term does contribute for $(1,0;1^{s/2-1})$: since $n_\vartriangle=p_\diamond=0$, we get $\epsilon=-1$. The result is
\begin{align}
\ket{\chi_{1,s}} \longleftrightarrow P^\alpha_{(1^{s/2})} -P^\alpha_{(1,0;1^{s/2-1})}
\end{align}
which does correspond to the formula given in \cite[Eq. (B.22)]{DLM_jhep}.

\section{Duality transformation}\label{sectionduality}

  Let $F_{r,s}(\alpha)$ be the superpolynomial representing the singular vector $|\chi_{r,s}\rangle$ in the R sector.  Here we show that there is a simple  
 duality formula involving $F_{r,s}(\alpha)$ which allows to go from the superpolynomial representation of  $|\chi_{r,s}\rangle$ to that of  $|\chi_{s,r}\rangle$.  This duality is slightly more complicated  than  in the NS sector \cite{DLM_jhep} and it will thus be worked out in detail.

The guiding observation in view of
 establishing the formula implementing the interchange of the Kac labels is the following relation
\begin{equation}
\eta_{r,s}(\alpha)=-\eta_{s,r}(1/\alpha)
\end{equation}
that follows from the parametrization given in \eqref{parametsing}.
This indicates that the label swapping is accompanied by the interchange of $\a\lrw1/\a$.
The operation that implements this interchange at the level of the sJacks is
  the following homomorphism naturally defined on power-sums (and whose action on sJacks is given below):
\begin{equation}\bar\omega_\alpha(p_n)=(-1)^{n-1}\alpha p_n,\qquad \bar\omega_\alpha(\tilde p_n)=(-1)^{n}\tilde p_n. 
\end{equation}
The last equation, {combined with the preservation of the commutation relations $[\d_n,p_m]=n\delta_{n,m}$ and  $\{ \ti\d_n,\ti p_m\}=\delta_{n,m}$,}
 imply that 
\begin{equation}
\bar \omega_\alpha(\partial_n)=(-1)^{n-1}\alpha^{-1}\partial_n,\qquad \bar \omega_\alpha(\tilde \partial_n)=(-1)^{n}\tilde \partial_n  .
\end{equation} 
Let us denote by 
$\mathcal{G}_1(\alpha,r,s)$ the expression for $\mathcal{G}_1$ in \eqref{GLpol}  evaluated with $\ga$ and $\eta$ replaced by their expressions in \eqref{parametsing}, and similarly for $\mathcal{L}_1$.  
Consider the action of $\bar \omega_\alpha$ on the first equation of  \eqref{2sv}:
\begin{equation}
 \bar\omega_\alpha\circ \mathcal{G}_1(\alpha,r,s)  \big(F_{r,s}(\a)\big) =\bar\omega_\alpha\circ \mathcal{G}_1(\alpha,r,s)\circ  \bar\omega_\alpha^{-1} \bar\omega_\alpha  \big(F_{r,s}(\a)\big)=0,
\end{equation}
and a similar equation with $\mathcal{G}$ replaced by $\mathcal{L}$.
We then determine by a direct calculation  how the two fundamental differential operators $ \mathcal{G}_1$ and $ \mathcal{L}_1$ -- given in \eqref{GLpol} -- do transform under a similarity transformation with $ \bar\omega_\alpha$. 
Using $\gamma(\alpha)=-\gamma(1/\alpha)$, we obtain  
\begin{equation}
 \bar\omega_\alpha\circ \mathcal{G}_1(\alpha,r,s) \circ   \bar\omega_\alpha^{-1} = \mathcal{G}_1(1/\alpha,s,r)  ,  \qquad \bar\omega_\alpha\circ \mathcal{L}_1(\alpha,r,s)\circ  \bar\omega_\alpha^{-1}=-\mathcal{L}_1(1/\alpha,s,r).
\end{equation}
{Therefore, the transformation  of $\mathcal {G}_1(\alpha,r,s)$ still reproduce $\mathcal{G}_1$ but with $\a$ changed into $1/\a$ and the Kac labels $r$ and $s$ interchanged. A similar result holds for $\mathcal{L}_1$, up to an overall sign which is irrelevant for the determination of singular vectors.} 
The transformed singular-vector relations should read
\beq
\mathcal{G}_1(1/\alpha,s,r)\big(F_{s,r}(1/\a)\big)=0,\qquad \mathcal{L}_1(1/\alpha,s,r)\big(F_{s,r}(1/\a)\big)=0.
\eeq
Therefore, we have the identification
\beq \bar\omega_\alpha \big(F_{r,s}(\a)\big)=F_{s,r}(1/\a)
\eeq
or equivalently, 
\beq
 \bar{\omega}_{1/\alpha}\big(F_{r,s}(1/\alpha)\big)=F_{s,r}(\a)
 \eeq
 which is the desired result. 
 
We have thus shown that, still assuming the parametrization given in \eqref{parametsing}, 
\begin{equation}
|\chi_{r,s}\rangle\quad \longleftrightarrow\quad F_{r,s}(\alpha)= \sum_{\Lambda\in \mathcal{A}_{r,s}}v_\Lambda^{(\alpha)}(r,s) P^{(\alpha)}_\La
\end{equation}
if and only if
\begin{equation}
|\chi_{s,r}\rangle\quad \longleftrightarrow\quad \bar{\omega}_{1/\alpha}\big(F_{r,s}(1/\alpha)\big)= \sum_{\Lambda\in \mathcal{A}_{r,s}}v_\Lambda^{(1/\alpha)}(r,s) \,\bar{\omega}_{1/\alpha}\big(P^{(1/\alpha)}_\La\big).
\end{equation}
This last expression can be made more explicit since
\begin{equation}
\tilde\omega_\alpha \big( P^{(\alpha)}_\La\big)=(-1)^{\binom{m}{2}}\alpha^m
\, \bar\omega_\alpha \big(  P^{(\alpha)}_\La \big), 
\end{equation}
where $\tilde\omega_\alpha$ is the homomorphism such that (cf. \cite[Eq. (4.30)]{DLM_jhep})
\begin{equation}
\tilde\omega_\alpha \big( P^{(\alpha)}_\La \big)=j_\La(\alpha)\,  P^{(1/\alpha)}_{\La'} . 
\end{equation}
The coefficient $j(\La)$ is the norm (up to a sign) of $P_\La^{(\a)}$ given by
\beq j_\La(\La)=\a^m\prod_{t\in\La}\frac{h^\up_\La(t)}{h^\dn_\La(t)}, 
\eeq
where $h^\dn_\La(t)$ and $h^\up_\La(t)$ are defined in \eqref{hdn} and \eqref{hup}.
 Therefore, we end up with the following expression representing $|\chi_{s,r}\rangle$:
\begin{equation}\label{swap}
|\chi_{s,r}\rangle\quad \longleftrightarrow\quad \sum_{\La\in \mathcal{A}_{r,s}}(-1)^{\binom{m}{2}} v_\Lambda^{(1/\alpha)}(r,s)\,\a^m 
\, j_\La(1/\alpha) \, P^{(\alpha)}_{\La'}.
\end{equation}

 Let us consider as an example, the construction of $\ket{\chi_{2,5}}$ out of $\ket{\chi_{5,2}}$, whose expression is
\begin{align}
	\ket{\chi_{5,2}} \longleftrightarrow P^{(\alpha)}_{{(5)}}+2\,{\frac { \left( 2+3\,\alpha \right)  \left( 5
\,\alpha+1 \right) P^{(\alpha)}_{{(4,1)}}}{ \left( 4\,\alpha+1 \right) 
 \left( 3\,\alpha+1 \right)  \left( 1+2\,\alpha \right) }}+2\,{\frac { \left( 5\,\alpha+1 \right) 
 \left( \alpha+2 \right) P^{(\alpha)}_{{(3,2)}}}{ \left( 3\,\alpha+1 \right)  \left( 1+2\,
\alpha \right) ^{2}}}\nonumber \\
 -4\,{
\frac { \left( \alpha+1 \right)  \left( 5\,\alpha+1 \right) P^{(\alpha)}_{{(4,1;)}}}{ \left( 4\,\alpha+1 \right)  \left( 3\,\alpha+1 \right) }}-2\,{\frac {
 \left( 2+3\,\alpha \right)  \left( 5\,\alpha+1 \right) P^{(\alpha)}_{{(3,2;)}}
}{ \left( 3\,\alpha+1 \right)  \left( 1+2\,\alpha \right) }}-P^{(\alpha)}_{{(
5,0;)}}.
\end{align}
According to \eqref{swap}, the singular vector $\ket{\chi_{2,5}}$ is given by
\begin{align}
&\left[ j_{(5)}P^{(1/\alpha)}_{{(5)^\prime}}+j_{(4,1)}{\frac { 2\left( 2+3\,\alpha \right)  \left( 5
\,\alpha+1 \right) P^{(1/\alpha)}_{{(4,1)^\prime}}}{ \left( 4\,\alpha+1 \right) 
 \left( 3\,\alpha+1 \right)  \left( 1+2\,\alpha \right) }}+j_{(3,2)}{\frac {2 \left( 5\,\alpha+1 \right) 
 \left( \alpha+2 \right) P^{(1/\alpha)}_{{(3,2)^\prime}}}{ \left( 3\,\alpha+1 \right)  \left( 1+2\,
\alpha \right) ^{2}}} \right. +\frac{(-1)^{\binom{2}{2}}}{\alpha^2} \nonumber \\ &\left.
 \times\left(-j_{(4,1;)}{
\frac {4 \left( \alpha+1 \right)  \left( 5\,\alpha+1 \right) P^{(1/\alpha)}_{{(4,1;)^\prime}}}{ \left( 4\,\alpha+1 \right)  \left( 3\,\alpha+1 \right) }}-j_{(3,2;)}{\frac {2
 \left( 2+3\,\alpha \right)  \left( 5\,\alpha+1 \right) P^{(1/\alpha)}_{{(3,2;)^\prime}}
}{ \left( 3\,\alpha+1 \right)  \left( 1+2\,\alpha \right) }}-j_{(5,0;)^\prime}P^{(1/\alpha)}_{{(
5,0;)^\prime}} \right) \right]_{\alpha \rightarrow \alpha^{-1}}.\label{exa}
\end{align}
The different $j_\Lambda$ are given by
\begin{align}
	\begin{array}{l l l }
	j_{(5)}=\frac {120{\alpha}^{5}}{ \left( 4\,\alpha+1 \right)  \left( 3\,
	\alpha+1 \right)  \left( 1+2\,\alpha \right)  \left( \alpha+1 \right)}
		&j_{(4,1)}=\frac {6 \left( 4\,\alpha+1 \right) {\alpha}^{4}}{ 	\left( 2+3\,\alpha \right)  \left( 1+2\,\alpha \right)  \left( \alpha+1 \right) }
			&j_{(3,2)}=\frac { \left( 3\,\alpha+1 \right)  \left( 1+2\,\alpha \right) {\alpha}^{3}}{ \left( \alpha+1 \right) ^{2} \left( \alpha+2 \right) }\\
	j_{(5,0;)}={\frac {24{\alpha}^{6}}{ \left( 4\,\alpha+1 \right)  \left( 3\,\alpha+1 \right)  \left( 1+2\,\alpha \right)  \left( \alpha+1 \right) }}
		&j_{(4,1;)}={\frac {{\alpha}^{5} \left( 4\,\alpha+1 \right) }{ \left( 1+2\,\alpha \right) ^{2} \left( \alpha+1 \right) ^{2}}}
 			&j_{(3,2;)}={\frac {{\alpha}^{4} \left( 3\,\alpha+1 \right) }{ \left( \alpha+1 \right) ^{2} \left( 2+3\,\alpha \right) }}.
	\end{array}
\end{align}
Substituting these expressions  into \eqref{exa}, and removing an overall factor in order to set the coefficient of  $P^{(\alpha)}_{{(2,2,1)}}$ equal to 1 yield 
\begin{align}
	P^{(\alpha)}_{{(2,2,1)}}+6\,{\frac { \left( \alpha+1 \right) P^{(\alpha)}_{{(2,1,1,1)}
}}{ \left( \alpha+3 \right)  \left( \alpha+2 \right) }}+60\,{\frac {
 \left( \alpha+1 \right) P^{(\alpha)}_{{(1,1,1,1,1)}}}{ \left( \alpha+5
 \right)  \left( \alpha+4 \right)  \left( \alpha+3 \right) }}+\alpha\,
P^{(\alpha)}_{{(1,0;2,2)}}\nonumber\\
+2\,{\frac { \left( \alpha+1 \right) \alpha\,P^{(\alpha)}_{{(1,0;2,1,1)}}}{ \left( \alpha+3 \right)  \left( \alpha+2 \right) }}+12\,
{\frac { \left( \alpha+1 \right) \alpha\,P^{(\alpha)}_{{(1,0;1,1,1,1)}}}{
 \left( \alpha+5 \right)  \left( \alpha+4 \right)  \left( \alpha+3
 \right) }}.
\end{align}
This agrees with the case $s=5$ of the general expression for the $|\chi_{2,s}\R$ singular vectors given in \cite[Eq. (B.23)]{DLM_jhep}.\\

More generally, we can recover the general expression for $|\chi_{2,s}\R$  (that is,  \cite[Eq. (B.23)]{DLM_jhep}) by applying the duality transformation to the  the closed-form expression for $|\chi_{s,2}\R$, but the analysis is somewhat technical and does not involve essential novelties.


\begin{appendix}

\section{Superpartitions,  superpolynomials and sJacks: a brief review}
\label{SsJ}

\subsection{Superpartitions} \label{A1}
A superpartition $\La$  is
a pair of partitions 
\begin{equation}\label{sppa}
\La=(\La^{a};\La^{s})=(\La_1,\ldots,\La_m;\La_{m+1},\ldots,\La_\ell),
\end{equation}
such that
\begin{equation}\label{sppb}
\La_1>\ldots>\La_m\geq0 \quad  \text{ and}
\quad \La_{m+1}\geq \La_{m+2} \geq \cdots \geq
\La_\ell > 0 \, .\end{equation}
\noindent 
The number $m$ is the fermionic degree of $\Lambda$ and  $\ell$ is its length. The bosonic degree is $|\La|=\sum_i\La_i$.
By removing the semi-coma and reordering the parts, we obtain an ordinary partition that we denote $\La^*$. The diagram of $\La$ is that of $\La^*$ with circles added to the rows corresponding to the parts of $\La^a$
 and ordered in length as if a circle was a half-box.
Finally, we will denote by $\La^\cd$ the partition obtained from the diagram of $\La$ by replacing circles by boxes. Here is an example, for which $\La=(4,3,0;4)$:
\beq \Lambda={\tableau[scY]{&&&&\bl\tcercle{}\\&&&\\&&&\bl\tcercle{}\\\bl\tcercle{} \\ }} 
 \quad \Longleftrightarrow \quad \Lambda^\circledast={\tableau[scY]{&&&&\\&&& \\&&&\\  \\ }} \quad  \Lambda^*={\tableau[scY]{&&&\\&&&\\ &&\\ \bl\\ }} .\eeq

  Given  a partition $\lambda$ (in our case, it is either $\La^*$ or $\La^\cd$), its
conjugate $\lambda'$ is the diagram
obtained by reflecting  $\lambda$ about the main diagonal.
In the main text, we make use of the following data: for a cell $t=(i,j)$ in $\lambda$, we define the arm and the leg of the box $t$ as
\begin{equation} \label{arml}
a_{\lambda}(t)=\lambda_i-j\, ,\qquad
l_{\lambda}(t)=\lambda_j'-i \, . 
\end{equation}
Note that conjugation is defined for superpartitions in the same way as for partitions: rows of $\La$ becomes columns of $\La'$. For instance 
\beq
\left(\;{\tableau[scY]{&&\bl\tcercle{}\\\bl\tcercle{} }}\right)'={\tableau[scY]{& \bl\tcercle{}\\ \\\bl\tcercle{} }}\eeq

\subsection{Superpolynomials}\label{A2}
Superpolynomials 
 polynomials in the usual commuting $N$ variables $x_1,\cdots ,x_N$  and the $N$ anticommuting variables $\ta_1,\cdots,\ta_N$.
Symmetric superpolynomials are invariant with respect to the interchange of $(x_i,\ta_i)\lrw (x_j,\ta_j)$ for any $i,j$ 
\cite{DLM1}. They are labelled by superpartitions.

The simplest example of a symmetric superpolynomial is the super-version of the monomial polynomials:\begin{equation}
m_\La(x,\theta)=\theta_{1}
\cdots\theta_{m} x_1^{\Lambda_1} \cdots x_N^{\Lambda_N}+\text{distinct permutations}
\end{equation} 
 {with the understanding that $\La_{\ell+1}=\cdots =\La_N=0$. }
An explicit example, for $N=4$ is
\begin{align}
m_{(1,0;1,1)}(x;\theta)=\;&\ta_1\ta_2(x_{1}-x_2)x_3x_4+\ta_1\ta_3(x_{1}-x_3)x_2x_4+\ta_1\ta_4(x_{1}-x_4)x_2x_3\nonumber\\+\,&\ta_2\ta_3(x_{2}-x_3)x_1x_4+\ta_2\ta_4(x_{2}-x_4)x_1x_3+\ta_3\ta_4(x_{3}-x_4)x_1x_2.
\end{align}

Another example is 
the super-power-sums\begin{equation}\label{spower}
p_\La=\tilde{p}_{\La_1}\cdots\tilde{p}_{\La_m}p_{\La_{m+1}}\cdots p_{\La_\ell},\qquad\text{where}\quad \tilde{p}_n=\sum_i\theta_ix_i^n\quad\text{and}\quad p_n=
\sum_ix_i^n \, .
\end{equation}  
Both $\{m_\La\}$ and $\{p_\La\}$ provide bases for the space of symmetric superpolynomials.

\subsection{Jack superpolynomials}\label{A3}
\label{charaphys}
The Jack superpolynomials (sJacks) $P_\La^{(\alpha)}$ can be characterized by the following
two conditions (e.g., see  \cite{DLMeva}). The first is {\it triangularity in the monomial basis:}
\beq \label{Ptriangular}
P_\La^{(\a)} =m_\La+\sum_{\Om<\La}
c_{\La\Om}(\alpha)\,m_\Om \, ,
\eeq
where $<$ refers to  the dominance order on
 superpartitions   \cite{DLMeva}:
\begin{equation} \label{eqorder1}
 \Omega\leq\Lambda \quad \text{iff}
 \quad \Omega^* \leq \Lambda^*\quad \text{and}\quad
\Omega^{\circledast} \leq  \Lambda^{\circledast} .
\end{equation}
We recall that the order on partitions is the usual dominance ordering:
\begin{equation}
\lambda \geq \mu \quad \iff \quad \sum_{i} \lambda_i =\sum_i \mu_i
\quad \text{and}\quad \lambda_1+ \cdots +\lambda_k \geq
\mu_1+ \cdots +\mu_k\, \quad \forall \; k \, .
\end{equation}
Pictorially,  $\Om<\La$ if $\Om$ can be obtained from $\La$ successively by moving down a box or a circle, which can be viewed as a sort of super-squeezing rule.
For example, 
\beq
{\tableau[scY]{&&&&\bl\tcercle{}}}
\quad > \quad
 {\tableau[scY]{&&\bl\tcercle{}\\ &}}\quad > \quad
 {\tableau[scY]{&\\&\\ \bl\tcercle{}}} \;.  \eeq

The second condition is {\it orthogonality in the power-sum basis:}
\begin{equation} \label{scap}
\LL \, 
{P_\La} \, | \, {P_\Om }\, \RR_\alpha=0\quad \text{if} \quad \La\ne \Om\, .
\eeq
The scalar product is defined as follows:
\begin{equation} \label{scap} \LL \, 
{p_\La} \, | \, {p_\Om }\, \RR_\alpha=(-1)^{\binom{m}2}\, \alpha^{{\ell}(\La)}\, z_{\La^s}
\delta_{\La,\Om}\,, \end{equation}
where $z_{\La^s} $ is given by
\begin{equation}  \label{zlam}
z_{\La^s}=\prod_{i \geq 1} i^{n_{\La^s}(i)} {n_{\La^s}(i)!}\, ,
\end{equation}
with $n_{\La^s}(i)$ the number of parts in $\La^s$ equal to $i$.

\end{appendix}


\begin{thebibliography}{99}
\addcontentsline{toc}{section}{References}

%

%

 \bibitem{AMOSa} H. Awata, Y. Matsuo, S. Odake and J. Shiraishi, {\it Collective field theory, Calogero-Sutherland model and generalized matrix models}, Phys.\ Lett.\ {\bf B347} (1995) 49--55;  {\it Excited states of the Calogero-Sutherland model and singular vectors of the $W_N$ algebra}, Nucl. Phys. {\bf B449} (1995) 347--374.
	
	
	
	

%

%


%
%

\bibitem{BBT}
A.A.\ Belavin, M.A., Bershtein, and G.M.\ Tarnopolsky,  \textit{Bases in coset conformal field theory from AGT correspondence and Macdonald polynomials at the roots of unity}, JHEP {\bf 03} (2013) 019 1--36.


\bibitem{BsA}
L.\ Benoit and Y.\ Saint-Aubin, {\it Singular vectors of the Neveu-Schwarz algebra},
Int.\ J.\ of Mod.\ Phys.\ A {\bf 7} (1992) 3023--3033.


\bibitem{BKT}
M. Bershadsky, V. Knizhnik, M. Teitelman, {\it Superconformal symmetry in two dimensions}
Phys.\ Lett.\, {\bf 151B} (1985) 31--36.








  \bibitem{DLM1}P.~Desrosiers, L.~Lapointe and P.~Mathieu, \emph{Supersymmetric Calogero-Moser-Sutherland models and Jack superpolynomials}, Nucl.\ Phys.\ {\bf B606} (2001)  547--582.










\bibitem{DLMeva} P.~Desrosiers, L.~Lapointe and P.~Mathieu, \emph{Evaluation and normalization of Jack polynomials in superspace},   Int.\ Math.\ Res.\ Not.\ (2012) 5267--5327.





\bibitem{DLM_jhep}
P.~Desrosiers, L.~Lapointe and P.~Mathieu,
{\it Superconformal field theory and Jack
superpolynomials}, JHEP 09 (2012) 037 1--41; slightly revised version available at arXiv:1205.0784v3.  


 \bibitem{CFT}
P.~Di Francesco, P. Mathieu, and D. S\'{e}n\'{e}chal,
{\em {Conformal Field Theory}},
Springer-Verlag, New York, 1997. 
 

\bibitem{Dorr} M.\  Dörrzapf, \textit{ Highest weight representations of the $N= 1$ Ramond algebra}, Nucl.\ Phys.\  {\bf B595} (2001) 605--653.




\bibitem{FQS}
D. Friedan, Z. Qiu, S.H. Shenker, {\it Superconformal invariance in two dimensions and the tricritical Ising model},
Phys.\ Rev.\ Lett.\  {\bf 151B} (1985) 37--43.














\bibitem{Mac} {I.~G.~ Macdonald},
                \emph{Symmetric functions and {H}all polynomials},
2nd ed., The Clarendon Press/Oxford University Press
                (1995).
                 
                 


\bibitem{MY}
 K. Mimachi and Y. Yamada, {\it Singular vectors of the Virasoro algebra in terms of Jack symmetric polynomials}, Commun.\ Math.\ Phys.\ {\bf174} (1995) 447--455.
 






\bibitem{SSAFR}
 R. Sakamoto, J. Shiraishi, D. Arnaudon, L. Frappat, and E. Ragoucy, {\it Correspondence between conformal field theory and Calogero-Sutherland model}, Nucl.\ Phys.\ {\bf B 704} (2005) 490--509. 
 

    

\bibitem{Uglov}
D. Uglov, {\it Yangian Gelfand-Zetlin bases, $gl_N$--Jack polynomials and computation of dynamical correlation functions in the spin Calogero-Sutherland model}, Commun. Math. Phys {\bf 191} (1998) 663-696;
{\it Symmetric functions and the Yangian decomposition of the Fock and Basic modules of the affine Lie algebra $\widehat {sl} (N)$}, arXiv  q-alg/9705010.



 


 



\bibitem{Watts} G.\ M.\ T.\  Watts, \textit{Null vectors of the superconformal algebra: The Ramond sector}, 
Nucl.\ Phys.\ B {\bf 407}  213--236.

 

\end{thebibliography}
\end{document}